# $Mn_3TeO_6$ – a new multiferroic material with two magnetic sub-structures

Li Zhao[1,3], Zhiwei Hu[1], Chang-Yang Kuo[1], Tun-Wen Pi[2], Maw-Kuen Wu[3], Liu Hao Tjeng[1], and Alexander C. Komarek*,[1]

[1] Max-Planck-Institute for Chemical Physics of Solids, Nöthnitzer Strasse 40, 01187 Dresden, Germany

[2] National Synchrotron Radiation Research Center, 101 Hsin-Ann Road, Hsinchu 30077, Taiwan

[3] Institute of Physics, Academia Sinica, 128 Sec. 2, Academia Road, Nankang, Taipei 11529, Taiwan

Abstract

From magnetic susceptibility, dielectric permittivity, electric polarization and specific heat measurements, we discover spin-induced ferroelectricity and magnetoelectric coupling in $Mn_3TeO_6$ and observe two successive magnetic transitions at low temperatures. A non-ferroelectric intermediate magnetic state occurs below 23 K and a multiferroic ground state emerges below 21 K. Moreover, $Mn_3TeO_6$ is a candidate for a multiferroic material where two types of incommensurate spin structures, cycloidal and helical, coexist. Theoretically, both spin substructures may contribute to the macro electric polarization via different mechanisms. This could open new ways of manipulating the ferroelectric polarization in a multiferroic material.

## 1 Introduction

Multiferroic materials have attracted enormous attention in the last decade because of their potential application in future devices [1–6]. The intrinsic cross-coupling between magnetic and electric properties opens up the possibility to manipulate the electric polarization by an external magnetic field and the magnetic structure by an applied electric field. The ferroelectric polarization in these materials can often be explained by the inverse Dzyaloshinskii–Moriya (DM) interaction $P_{ij} = A\hat{e}_{ij} \times (S_i \times S_j)$ [7–9] or by the $P_{ij} \propto (S_i \cdot \hat{e}_{ij})S_i - (S_j \cdot \hat{e}_{ij})S_j$ term in the Arima model [10, 11]. Most multiferroic materials are based on oxides. Several halide and chalcogenide compounds [12–14] are also known to exhibit multiferroicity.

Very recently, the transition metal orthotellurates $M_3TeO_6$ (M = $Mn^{2+}$, $Co^{2+}$, and $Ni^{2+}$) with a corundum-related structure have attracted considerable attention [15–21]. A magnetic field induced spontaneous electric polarization has been discovered in $Co_3TeO_6$ [15] and a colossal magnetoelectric (CME) effect below its AFM ordering temperature of about 52 K was reported in $Ni_3TeO_6$ [19]. Here we focus on $Mn_3TeO_6$ which was found to exhibit a complex spin structure with both helical and cycloidal components [20, 21]. We have carried out a comprehensive study of the magnetic susceptibility, specific heat, local electronic structure, dielectric permittivity, and electric polarization, and discovered that $Mn_3TeO_6$ is multiferroic. The coexistence of a helical and of a neighboring cycloidal spin structure in the ferroelectric phase makes this multiferroic material unique among other multiferroic materials.

## 2 Experimental section

Polycrystalline $Mn_3TeO_6$ samples were prepared by solid-state reaction. The mixture of high-purity $MnO_2$ and $TeO_2$ in the stoichiometric ratio of 3:1 was thoroughly ground and pressed into pellets. Subsequently it was sintered in a tubular furnace with an Ar flow at 750–800 ° for 72 hours with several intermediate grindings. The light brown bulk samples have been characterized by X-ray powder diffraction, confirming that our samples are single phase. Magnetic properties were measured in a SQUID magnetometer (MPMS-5XL, Quantum Design).

To measure the dielectric properties of $Mn_3TeO_6$, the polycrystalline samples were polished to thin plates with thickness of 0.2–0.5 mm. Silver paint was applied to both sides as electrodes to form parallel plate capacitors whose capacitance are proportional to the dielectric constant ($\varepsilon$). The samples are glued on the cryogenic stage of a homemade probe inserted into a Quantum Design 9T PPMS. A high-precision capacitance bridge (AH 2700A, Andeen-Hagerling, Inc.) was used for dielectric measurement. We tried various excitation levels (from 1 V to 15 V) and sweeping rates, and no apparent difference was found in the different measuring conditions.

The electric polarization has been obtained by measurements of the pyroelectric current. Firstly we polarized the specimens with a static electric field of 300–800 kV/m during the cooling process, then removed the electric field and short-circuited both sides of the sample at low temperature for about one hour to remove the possible trapped interfacial charge carriers. The pyroelectric current was measured during the warming process at a heating rate of about 3 K/min. The electric polarization ($P$) has been obtained from the integration of the measured pyroelectric current.

Furthermore, soft X-ray absorption spectroscopy (XAS) at the Mn-$L_{2,3}$ edge has been performed in the total electron yield mode with a photon energy resolution of 0.2 eV at the BL08B beamline of National Synchrotron Radiation Research Centre in Taiwan. Clean sample surfaces were obtained by fracturing the samples in situ just before collecting the data in an ultrahigh vacuum chamber with the pressure below $10^{-9}$ mbar. The Mn-$L_{2,3}$ XAS spectrum of a single crystal of MnO was measured simultaneously as a standard reference.

## 3 Results and discussion

We have synthesized single phase polycrystalline $Mn_3TeO_6$ samples by solid-state reaction. Details can be found in Section 2. The temperature dependence of the magnetic susceptibility ($\chi$) of $Mn_3TeO_6$ measured in the field of $H$ = 1000 Oe is shown in Fig. 1a. The zero-field-cooling (ZFC) and field-cooling (FC) curves exhibit the typical behavior for an antiferromagnet. The Curie–Weiss temperature $\theta_{CW}$ from a fit to $\chi(T) = C/(T-\theta_{CW})$ amounts to –112 K, indicating that the interactions in Mn$_3$TeO$_6$ are antiferromagnetic (AFM).

Furthermore, the effective magnetic moment $\mu_{eff}$ amounts to 5.8 $\mu_B$, indicating that the Mn ion is divalent with the $S = 5/2$ spin state. For temperatures below about 130 K, $\chi(T)$ deviates clearly from the Curie–Weiss behavior, suggesting that short-range correlations start to develop. At ~ 23 K we observe a sharp kink in the susceptibility (see also the sharp peak in d$\chi$/d$T(T)$ as shown in the inset of Fig. 1a. We can identify this temperature as the Néel temperature of Mn$_3$TeO$_6$, since the temperature dependent specific heat as displayed in Fig. 1b shows a sharp $\lambda$-shaped peak indicating the release of substantial amount of entropy. Here we will use the notation $T_{N1}$ for this 23 K Néel temperature to differentiate it from a second spin transition at a lower temperature $T_{N2}$ described later. We note that $T_{N1}$ is much lower than $\theta_{CW}$. The $\theta_{CW}/T_{N1}$ ratio is about ~ 4.9, indicating considerable magnetic frustration in Mn$_3$TeO$_6$ [22].

To understand better the origin of the 5.8$\mu_B$ effective magnetic moment, we have carried soft X-ray absorption experiments at the Mn $L_{2,3}$ edge. Figure 1c shows the spectrum of Mn$_3$TeO$_6$ together with those of MnO and LaMnO$_3$ as reference for Mn$^{2+}$ and Mn$^{3+}$ ions, respectively. The features and their energy positions of the Mn$_3$TeO$_6$ spectrum resemble very much those of the MnO but are very different from the ones of LaMnO$_3$. We can therefore safely conclude that the Mn ion in Mn$_3$TeO$_6$ is divalent and in the high spin ($S = 5/2$) state [23].

Figures 2a,b show the temperature-dependence of the dielectric constant ($\varepsilon$) and the dielectric loss (tan$\delta$) measured in zero field using different measurement frequencies (ranging from 1 kHz to 20 kHz). The most apparent anomaly in $\varepsilon$ and tan$\delta$ is a sharp $\lambda$-like peak at $T_{N2}$ ~ 21 K, which is indicative for the emergence of a ferroelectric transition in Mn$_3$TeO$_6$ at $T_{N2}$. The frequency independence of this sharp peak excludes possible spurious artefacts arising from grain boundaries, contact etc. We have performed complementary pyroelectric measurements. As shown in Fig. 2c a non-zero electric polarization ($P$) develops below $T_{N2}$ and saturates at a value of about 2.7 $\mu$C/m$^2$. The observed polarization can be inverted with opposite poling of the applied electric field unambiguously demonstrating the ferroelectric nature of the transition at $T_{N2}$. Hence, Mn$_3$TeO$_6$ is a multiferroic material. The size of $P$ is comparable to that of many other known multiferroics in which the electric polarization is induced by spin structures at low temperatures [5, 6].

In addition to the sharp peak, a kink-like feature also appears at $T_{N1}$ in $\varepsilon(T)$, whereas no corresponding anomalies can be observed in tan$\delta(T)$. This demonstrates that Mn$_3$TeO$_6$ is not ferroelectric between $T_{N1}$ and $T_{N2}$. An anomalous drop in $\varepsilon(T)$ at the Néel temperature (denoted as $T_{N1}$ here) has been observed also in other non-polar AFM systems [24, 25]. The occurrence of an electric polarization at a temperature lower than the Néel temperature is often observed in multiferroic materials with an intrinsic coupling between dielectric and magnetic properties [26–28]. Hence, Mn$_3$TeO$_6$ is indeed a new magnetically driven multiferroic material in which a sizeable magnetoelectric coupling can be expected due to the close coupling between magnetism and ferroelectric properties.

The magnetoelectric coupling effects in Mn$_3$TeO$_6$ were measured as a function of external magnetic field ($H$ = 0–9 T) and temperature. Figure 2d shows the magnetodielectric data at several temperatures below and above the magnetic transition measured at 1 kHz. Here, the magnetodielectric effect is characterized by the coefficient, $[\varepsilon(H)-\varepsilon(H=0)]/\varepsilon(H = 0)$. For $T = 25$ K above $T_{N1}$, there is almost no field-induced change in $\varepsilon$. Below $T_{N1}$ the magnetodielectric coefficient increases quadratically with $H$ within the low field region – which is consistent with the empirical analysis based on Landau theory [24] – and tends to saturate at higher fields. In order to investigate the magnetic field effects in $Mn_3TeO_6$ systematically, we plot $\varepsilon$ as function of $T$ under different external fields ($H = 0$–9 T). Figure 2e shows that between $T_{N2}$ and $T_{N1}$, the $\varepsilon(T)$ curves are almost unaffected by $H$. Around $T_{N1}$ there are not any discernable anomalies in dielectric loss up to the highest fields $H = 9$ T indicating no field induced FE above $T_{N2}$. But around $T_{N2}$, the dielectric peak is suppressed by $H$ and is shifted to lower temperatures with increasing field. The field suppression of ferroelectricity in $Mn_3TeO_6$ can be also observed within electric polarization measurements, see Fig. 2f. At 5 K, the saturated $P$ decreases gradually from 2.8 $\mu C/m^2$ at zero field to 1.9 $\mu C/m^2$ at 9 T. The ferroelectric transition temperature $T_{N2}$ deduced from the peak position in $\varepsilon(T)$ decreases continuously from ~ 21 K for $H = 0$ T to ~ 19.8 K for $H = 9$ T. The strong field-dependent behavior of $\varepsilon(T)$ and P indicates the magnetic origin of ferroelectricity in $Mn_3TeO_6$.

Our results reveal two successive phase transitions in $Mn_3TeO_6$. This can also be seen in the specific heat measurements. Close to the large lambda-peak at $T_{N1}$ we can detect a weak hump-like anomaly of $C_p/T$ around the ferroelectric transition $T_{N2}$ (marked with the vertical dashed line in Fig. 3a.

We now can draw a magnetic phase diagram of $Mn_3TeO_6$ as shown in Fig. 3. The two successive magnetic phase transitions found in $Mn_3TeO_6$ resemble on the observations in other magnetically driven multiferroic materials with spiral spin structures, e.g. in $LiCu_2O_2$ (~ 24 K and ~22 K) [26], $TbMnO_3$ (~41 K and ~28 K) [1] and $NaFeSi_2O_6$ (~ 9 K and ~ 6 K) [28]. These materials first undergo a transition from a paramagnetic state to a sinusoidal SDW with collinear spin structure, which does not give rise to any ferroelectricity. As temperature decreases further, the order parameter grows and the systems undergo a second transition from SDW to the spiral state [27]. Similar to these known multiferroic materials, the FE polarization in $Mn_3TeO_6$ occurs only below a second transition temperature $T_{N2}$, which is slightly lower than the first AFM transition temperature $T_{N1}$. Remarkably, the spin structure in the ground state of $Mn_3TeO_6$, reported by Ivanov et al. [20] is much more complex than in other known multiferroic materials. In $Mn_3TeO_6$ the AFM state hosts two sets of incommensurate non-collinear spin configurations of fundamentally different types – one is a helical spin structure and the other one is basically a cycloidal one. Below $T_{N2}$, both helical and cycloidal spin sub-structures in $Mn_3TeO_6$ are theoretically able to induce macroscopic electric polarization, but via different mechanisms. The cycloidal-like spin structure hosted in the Mn(2) chains can produce an electric polarization $P$ perpendicular to $c$-axis via the inverse DM interaction, while the 'screw-type' helical spin sub-structure hosted within the Mn(1) chains is expected to induce a polarization along the $c$-axis according to the Arima model. Studies of the interplay between both spin substructures in single crystal $Mn_3TeO_6$ would be highly interesting since the presence of two different kind of coexisting spiral spin configurations – helical and cycloidal – is an unprecedented situation in a multiferroic material.

**4 Conclusions**

To summarize, in this work, we presented dielectric and pyroelectric measurements with high resolution on our bulk polycrystalline $Mn_3TeO_6$ samples. For the first time, we observe strong magnetoelectric coupling and spin-induced ferroelectricity in $Mn_3TeO_6$. Hence, $Mn_3TeO_6$ is a new multiferroic material with complex magnetic structure and with promising magnetoelectric properties. We observed successive two-step magnetic transitions: a non-FE intermediate magnetic state occurs below 23 K and a multiferroic ground state emerges below 21 K. In $Mn_3TeO_6$. Each of the two spin sub-structures might be able to produce macro electric polarization - either via inverse DM interaction for the cycloidal spin sub-structure or via the Arima mechanism for the helical spin sub-structure. The presence of two spin sub-structures might open new ways of manipulating the ferroelectric polarization in a multiferroic material.

**Acknowledgements**

We would like to thank H. Borrmann and his team for powder X-ray diffraction measurements and Ch. Becker,T. Mende and S. Wirth for their support on the construction of the dielectric measurement devices at MPI CPfS. We also acknowledge Z.W. Li, A. Keil and H.J. Guo for their help.

Figures

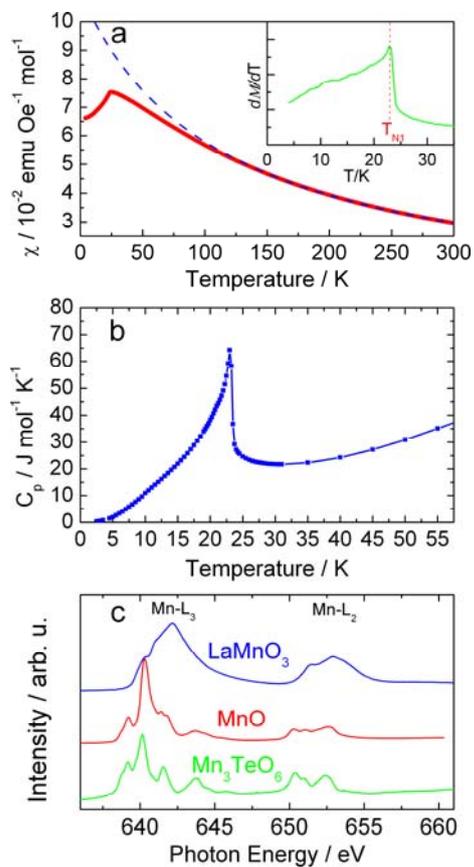

**Figure 1.** (a) The temperature dependence of magnetic susceptibility ($\chi$) of $Mn_3TeO_6$. The dashed line is the Curie-Weiss fit. The temperature derivative ($d\chi/dT$) is shown in the inset with a vertical dashed line marking its peak. (b) The temperature dependence of the specific heat of $Mn_3TeO_6$. (c) The Mn-$L_{2,3}$ XAS spectra of $Mn_3TeO_6$ and of MnO and $LaMnO_3$ as a $Mn^{2+}$ and $Mn^{3+}$ references, respectively).

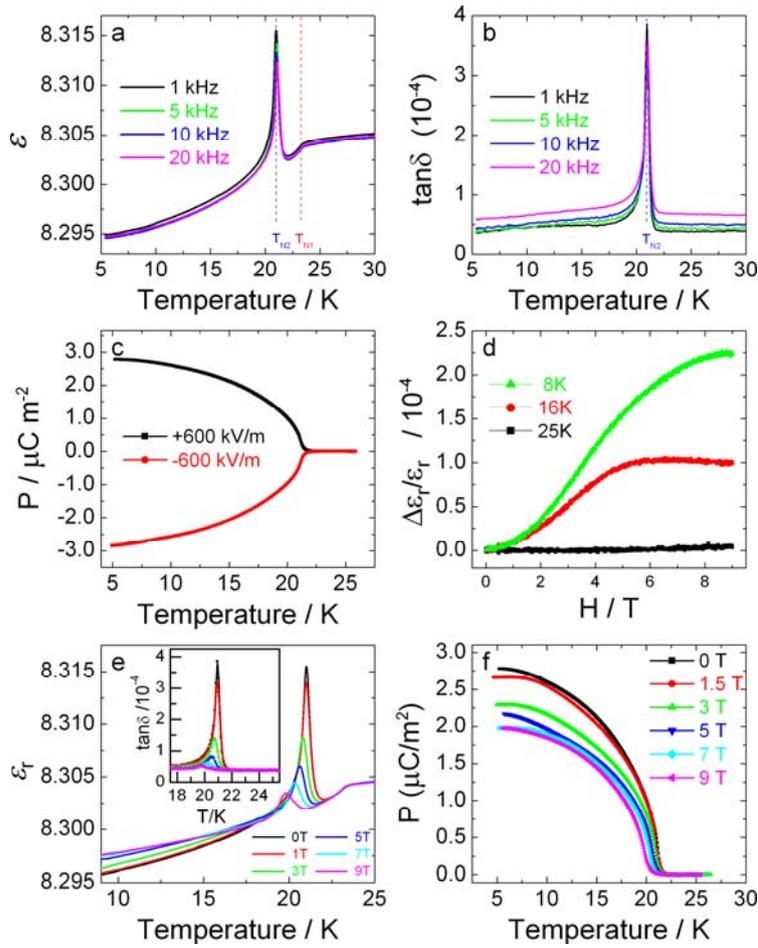

**Figure 2.** (a) $\varepsilon(T)$ of $Mn_3TeO_6$ measured at different frequencies in zero field. (b) The corresponding dielectric loss (tan$\delta$). (c) The temperature dependent electric polarization in zero field, measured with both positive and negative poling electric fields applied. (d) The magnetic field induced change of the dielectric constant (measured at 1 kHz) at different temperatures (T=8, 16, 25 K). (e) $\varepsilon(T)$ of $Mn_3TeO_6$ measured at 1 kHz in different magnetic field (H=0, 1, 3, 5, 7, 9 T). the corresponding dielelctric loss is shown in the inset. (f) the electric polarization $P$. measured in different H (0- 9 T).

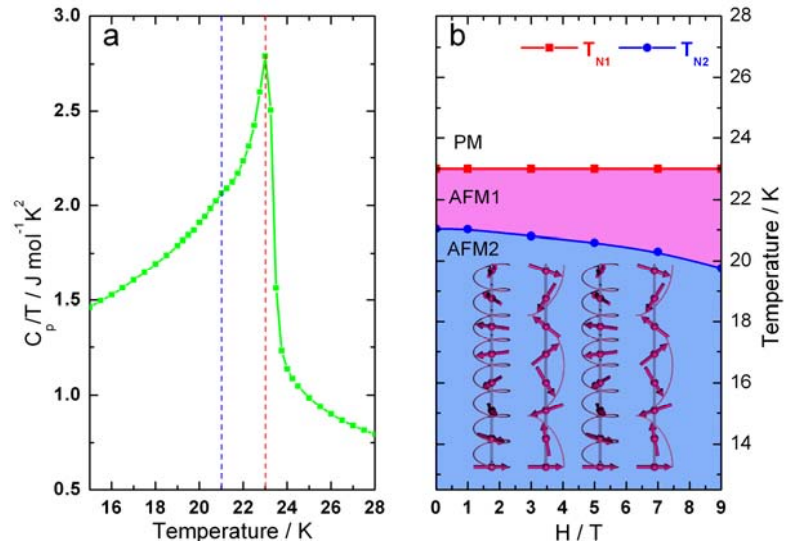

**Figure 3.** (a) The temperature dependence of C$_p$(T)/T. The two dashed lines indicate the anomalies at T$_{N1}$ and T$_{N2}$. (b) The magnetic phase diagram of Mn3TeO6 with the schematic drawing of the two spin sub-structures occurring below T$_{N2}$.